\documentclass[10pt,a4paper,twoside]{article}
\usepackage{graphicx,hyperref}
\pdfoutput=1
\usepackage{bm}
\usepackage{ulem}
\usepackage{tikz}
\usepackage{amsmath}
\usepackage{amssymb,latexsym,amsfonts,booktabs}

\usepackage{color}

\title{New activity pattern\\ in human interactive dynamics}

\author{Marco Formentin\thanks{Dipartimento di Fisica e Astronomia ``G. Galilei'', Universit\`a di\
Padova, via Marzolo 8, I-35151 Padova (Italy); e-mail addresses: marco.formentin@rub.de, maritan@pd.infn.it} \and Alberto Lovison \thanks{Dipartimento di Matematica,
Universit\`a di Padova, via Trieste 63, I-35121 Padova (Italy); e-mail address: lovison@math.unipd.it}\and Amos Maritan\footnotemark[1]\ \and Giovanni Zanzotto\thanks{Dipartimento di Psicologia Generale,  Universit\`a di Padova, via Venezia 12, I-35131 Padova (Italy); e-mail address: zanzotto@dmsa.unipd.it}}

\begin{document}

\maketitle

\begin{abstract}
We investigate the response function of human agents as demonstrated by written correspondence, uncovering a new universal pattern for how the reactive dynamics of individuals is distributed across the set of each agent's contacts. In long-term empirical data on email, we find that the set of response times considered separately for the messages to each different correspondent of a given writer, generate a family of heavy-tailed distributions, which have largely the same features for all agents, and whose characteristic times grow exponentially with the rank of each correspondent. We furthermore show that this universal behavioral pattern emerges robustly by considering weighted moving averages of the priority-conditioned response-time probabilities generated by a basic prioritization model. Our findings clarify how the range of priorities in the inputs from one's environment underpin and shape the dynamics of agents embedded in a net of reactive relations. These newly revealed activity patterns might be present in other general interactive environments, and constrain future models of communication and interaction networks, affecting their architecture and evolution. 
\end{abstract}

\vspace{0.3cm}


\section{Introduction}

The interaction dynamics of animal and human agents is of interest in many theoretical and applied domains of science, from ecology to sociology to economics. Especially interesting is the clarification of the response function of humans, which has been investigated in a variety of contexts \cite{Henderson:2001yq,Wang:2010vn,Dezso:2006kx,Goncalves:2008yq,Gao:2010fj,Vazquez:2006zr},  a paradigmatic case being written correspondence, especially through email.

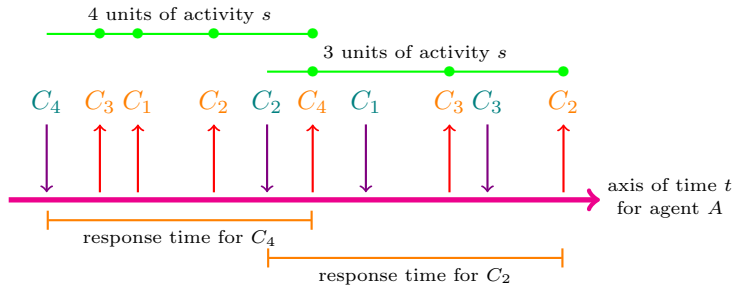
\begin{figure}
\centering
\begin{tikzpicture} 
\draw [->,magenta,line width=2pt] (-0.2,1) -- (7.58,1);

\coordinate [label={[black]center: $\mbox{\scriptsize{axis of time $t$}}$}] (C) at (8.5,1.2);
\coordinate [label={[black]center: $\mbox{\scriptsize{for agent $A$}}$}] (L) at (8.5,0.88);

\draw [->,violet,thick] (0.3,2) -- (0.3,1.1);
\coordinate [label={[teal]center: $C_4$}] (C) at (0.3,2.3);
\draw [<-,red,thick] (1,2) -- (1,1.1);
\coordinate [label={[orange]center: $C_3$}] (C) at (1,2.3);
\draw [<-,red,thick] (1.5,2) -- (1.5,1.1);
\coordinate [label={[orange]center: $C_1$}] (C) at (1.5,2.3);
\draw [<-,red,thick] (2.5,2) -- (2.5,1.1);
\coordinate [label={[orange]center: $C_2$}] (C) at (2.5,2.3);
\draw [->,violet,thick] (3.2,2) -- (3.2,1.1);
\coordinate [label={[teal]center: $C_2$}] (C) at (3.2,2.3);
\draw [<-,red,thick] (3.8,2) -- (3.8,1.1);
\coordinate [label={[orange]center: $C_4$}] (C) at (3.8,2.3);
\draw [->,violet,thick] (4.5,2) -- (4.5,1.1);
\coordinate [label={[teal]center: $C_1$}] (C) at (4.5,2.3);
\draw [<-,red,thick] (5.6,2) -- (5.6,1.1);
\coordinate [label={[orange]center: $C_3$}] (C) at (5.6,2.3);
\draw [->,violet,thick] (6.1,2) -- (6.1 ,1.1);
\coordinate [label={[teal]center: $C_3$}] (C) at (6.1,2.3);
\draw [<-,red,thick] (7.1,2) -- (7.1,1.1);
\coordinate [label={[orange]center: $C_2$}] (C) at (7.1,2.3);

\draw [|-|,orange,thick] (0.3,0.73) -- (3.8,0.73) node [black,pos=0.5, below] {\scriptsize{{response time for $C_4$ }}};

\draw [-,green,thick] (0.3,3.2) -- (3.8,3.2) node [black,pos=0.5, above] {\scriptsize{{4 units of activity  $s$}}};

\coordinate [label={[green]center: $\bullet$}] (C) at (1,3.2); \coordinate [label={[green]center: $\bullet$}] (C) at (1.5,3.2);
\coordinate [label={[green]center: $\bullet$}] (C) at (2.5,3.2);\coordinate [label={[green]center: $\bullet$}] (C) at (3.8,3.2);

\draw [|-|,orange,thick] (3.2,0.23) -- (7.1,0.23) node [black,pos=0.5, below] {\scriptsize{{response time for $C_2$ }}};

\draw [-,green,thick] (3.2,2.7) -- (7.1,2.7) node [black,pos=0.5, above] {\scriptsize{{3 units of activity  $s$}}};

\coordinate [label={[green]center: $\bullet$}] (C) at (3.8,2.7); 
\coordinate [label={[green]center: $\bullet$}] (C) at (5.6,2.7);\coordinate [label={[green]center: $\bullet$}] (C) at (7.1,2.7);

\end{tikzpicture}
\caption{\footnotesize Activity clock for a node in an interaction network. Representation of the node's temporal activity, in this case written communication, along the axis of time $t$ for an agent $A$, typically measured in seconds for email data. Arrows pointing into the $t$ axis mark incoming messages from the indicated correspondents $C_1$, $C_2$, $\dots$, of $A$. Arrows pointing out of the $t$ axis mark response messages from $A$ to the same correspondents. The intervals between the outgoing arrows define the inter-event times of $A$. The response times (RTs) of $A$ pertaining to each correspondent $C_i$ can either be clocked through time $t$ (in brown), or (in green) through the activity parameter $s$ which counts the number of outgoing messages from $A$. See the text for definitions.}
\label{cartoon}
\end{figure}

When each person is viewed as the node of a graph, written communication generates an evolving weighted and directed network whose large-scale structure and dynamics are still virtually unknown. Many interesting facts have emerged from the investigation of a number of email or paper mail databases collecting basic empirical information on written correspondence spanning from a few months  \cite{Eckmann:2004vn, WuZhou2010} to several decades of writers' activity \cite{Vazquez:2006zr, Oliveira:2005fk, quwang2011, Malmgren:2009rt, UWC}. Intermittency was observed in the dynamics of correspondence writers, with bursts of events separated by long pauses, with non-Poissonian, heavy-tailed statistics in both the agents' inter-event times and response times (RTs), see the definitions below and Fig.~\ref{cartoon}. This also relates to the heavy-tailed temporal distributions observed in human and animal behavior and locomotion \cite{Henderson:2001yq, Wang:2010vn,Dezso:2006kx,Goncalves:2008yq,Gao:2010fj, Vazquez:2006zr, HANAI:kx, crane2010, topiAmos, foraging, virtual_IET_2015}.

A number of approaches have been used to characterize the features of the empirical time statistics of written communication \cite{Vazquez:2006zr, Barabasi:2005yq, Grinstein2008, IAT1,IAT2, blanchard2007, Malmgren:2008ys, min2009, cho2010, kim2010,  jo2011}, with debated indications of scaling behavior for the waiting times, and for their possible modeling through priority queueing.
A new method for the analysis of these human reactive phenomena has been recently proposed \cite{UWC}, through which it was shown that, in particular, the mechanisms underpinning the response-time (RT) statistics of written correspondence are best understood, rather than in terms of standard time $t$, in terms of an agent's activity, i.e.~by a 'proper time' parameter $s \in \mathbb{N}^+$ counting an agent's outbound messages. This approach \cite{UWC} disentangles from the overall time dynamics of writers the contributions due to their spontaneous pauses between messages, and helped uncover universal power-law features in the RT statistics on written correspondence when the $s$-clocking is utilized, rather than the usual $t$-clocking (Fig.~\ref{cartoon}).

Despite the insight given by such earlier enquiries, information of primary importance about the basic features of human interaction is still lacking, both in the data analysis and modeling. First and foremost, solely the total RT distribution $P(\sigma)$ of correspondence writers (with $\sigma = \Delta s$) has so far been considered in the literature, and it is presently unknown in which way the overall interaction of a given agent $A$ is distributed among all of her distinct targets. For instance, the response statistics of a writer $A$ separately with each one of her correspondents have so far never been obtained. This lack of empirical analysis parallels the fact that some main aspects of priority modeling have also remained unexplored in the above context. 

In the present study we go beyond the analysis of the total RT distribution $P(\sigma)$ of correspondence writers, and investigate how the writers' activity depends on the identity of their distinct contacts. As was done in Ref.~[28], where the voice-call inter-event times of cell-phone users with their distinct contacts have been considered, such an analysis is the first, natural step in a more in-depth investigation of the activity patterns in interaction networks, and sheds light on their structure that cannot be obtained from the sole total distribution $P(\sigma)$ of the involved agents. We utilize for our inquiry the database presented in Ref.~[13] 
and briefly described below, which is the most complete long-term email dataset currently available in the literature. Our findings reveal a new universal behavioral patterns as well as new modeling effects, evidencing hitherto unknown universal aspects of human interaction. The analysis also illustrates how priority, which in the model is a hidden variable not immediately linked to real data, operates in the generation of a dynamics in accord with empirical observations.

\section{Database and definitions.~Proper time}
Our written communication data concern the full server-recorded activity of all the email accounts belonging to a Department of a large EU university during two years (see also Ref.~[13]). 
The collected data are in the form {\{\sc sender, receiver, timestamp\}}, with senders and receivers conventionally numbered for identification, and timestamps given in seconds.

Referring to an agent $A$, the response times (RTs) are defined as the time intervals $\tau = \Delta t$ (in seconds) separating the arrival of any message $\mathcal{M}$ from any agent $B$ to $A$, and the first ensuing message $\mathcal{M}$$'$ going from $A$ to $B$, independently of the subject or contents of $\mathcal{M}$ or $\mathcal{M}$$'$. Following Ref.~[13],
to extricate from the time dynamics of $A$ the contributions due to $A$'s pauses between messages (given by the individual inter-event time distribution $P_{I}(\tau)$ of $A$), we introduce the activity parameter (proper time) $s \in \mathbb{N}^+$ of $A$, which clocks the number of outgoing messages from $A$. The RTs of $A$ are thus defined by counting the number $\sigma = \Delta s$ of outgoing messages from $A$ intervening between the same messages $\mathcal{M}$ and $\mathcal{M}'$ as above, as represented in Fig.~\ref{cartoon}.

Out of all the nominal monitored accounts in our dataset, we have analyzed the 300 most active agents, whose activity comprises from a minimum of 390 to $\sim$$10^4$ total RTs. A large percentage of these 300 writers have in the order of a few thousand RTs, distributed over a number $n$ of distinct correspondents ranging from less than $100$ to almost $1000$. The \href{https://www.researchgate.net/publication/280645729_Supplementary_figure_for_New_activity_pattern_in_human_interactive_dynamics}{Supplementary Figure}
\footnote{\tiny{The Supplementary Figure ($\sim$ 9 MB) is available at\\ https://www.researchgate.net/publication/280645729\_Supplementary\_figure\_for\_New\_activity\_pattern\_in\_human\_interactive\_dynamics}}
shows explicitly the RT statistics pertaining to the 84 most active, and the 12 least active, among such 300 agents. 

\begin{figure}[t!]\centering
 \includegraphics[width=70mm,clip=]{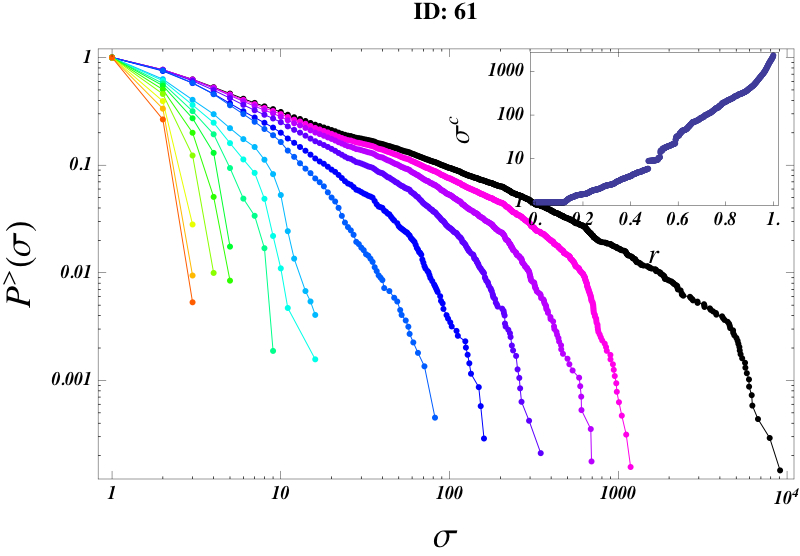} 
\caption{\footnotesize Empirical behavior of a human agent. Log-log plots of the empirical response-time (RT) inverse cumulative distributions  $P_{i}^{>}(\sigma)$ in the family $\mathcal{F}_{A}$ of \eqref{friends}, for a typical email writer $A$ in the database. 
For clarity, only some of the $n$ curves $P_{i}^{>}(\sigma)$ are plotted, for geometrically growing values of the normalized correspondent rank $r = \frac{i}{n} \in [0, 1]$ from left to right.  The distribution $P^{>}_1(\sigma)$ is the lowest curve in red, while the total RT distribution $P^{>}_{n}(\sigma) = P^{>}(\sigma)$ is the upper-most curve in black. The inset, whose horizontal axes report the normalized rank $r$, shows the linear-log plot of the values $\sigma_{i}^{c} = \frac{<\sigma^2>}{<\sigma>}$ for the distributions $P_{i}^{>}(\sigma)$. This dictates how the curves $P_{i}^{>}(\sigma)$ progressively extend to the right over the $\sigma$-axis for growing $i$, reaching the black total cumulative distribution $P^{>}(\sigma) = P^{>}_{n}(\sigma)$ of $A$ for $i = n$. The characteristic times $\sigma_{i}^{c}$ in the inset grow roughly exponentially as $r \uparrow 1$, see also Figs.~\ref{friends_empirical}-\ref{typical}. More statistics are given in the \href{https://www.researchgate.net/publication/280645729_Supplementary_figure_for_New_activity_pattern_in_human_interactive_dynamics}{Supplementary Figure}.
}
\label{one_agent}
\end{figure}

\section{Empirical results:~new universal behavioral pattern}~
We break down the activity of a writer $A$ by considering the RTs of $A$ separately for each one of her correspondents $C_i$ (see Fig.~\ref{cartoon}). We then rank $A$'s correspondents $C_1$,  $C_2$, $\dots$, through their growing characteristic ($s$-)times $\tilde{\sigma}_{i}^{c} =  \frac{<\sigma^2>}{<\sigma>}$, the latter being computed from the set of $s$-clocked RTs that $A$ generates with each distinct $C_i$. Then, from the empirical data regarding $A$, we obtain the ordered family of $s$-clocked distributions

\begin{equation}
\mathcal{F}_{A} = \{P^{>}_{i}   (\sigma), 
\;\;  i = 1, 2, ..., n   \;\},
\label{friends}
\end{equation}
where $n$ is the total number of $A$'s correspondents. 
\begin{figure}[t!] \centering
\begin{tabular}{ccc}
\includegraphics[width=40mm,clip=]{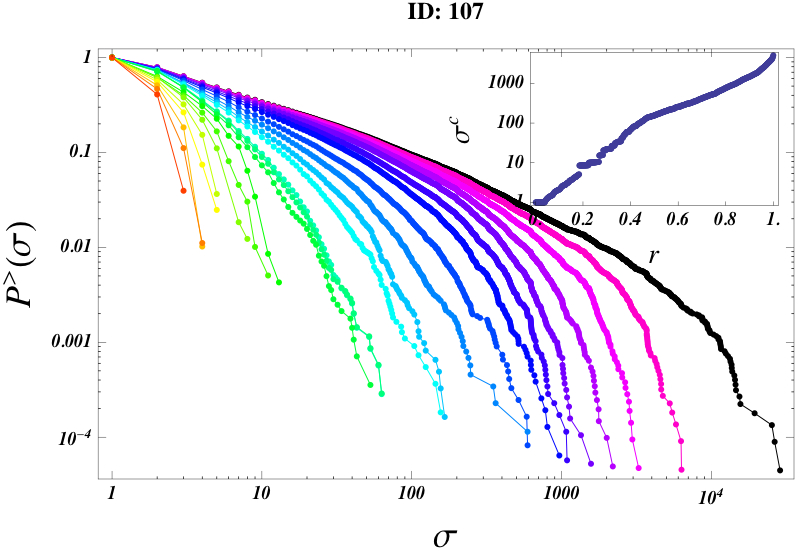} &\hspace{1mm} &  \includegraphics[width=40mm,clip=]{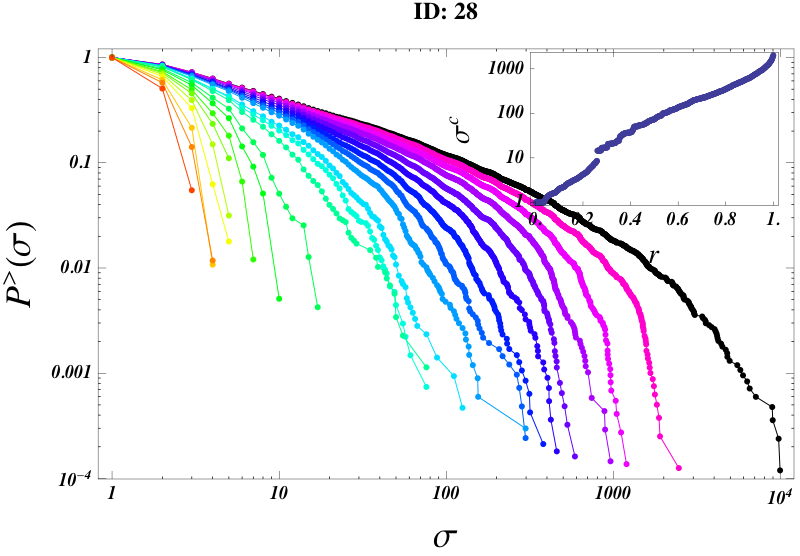}\\
\includegraphics[width=40mm,clip=]{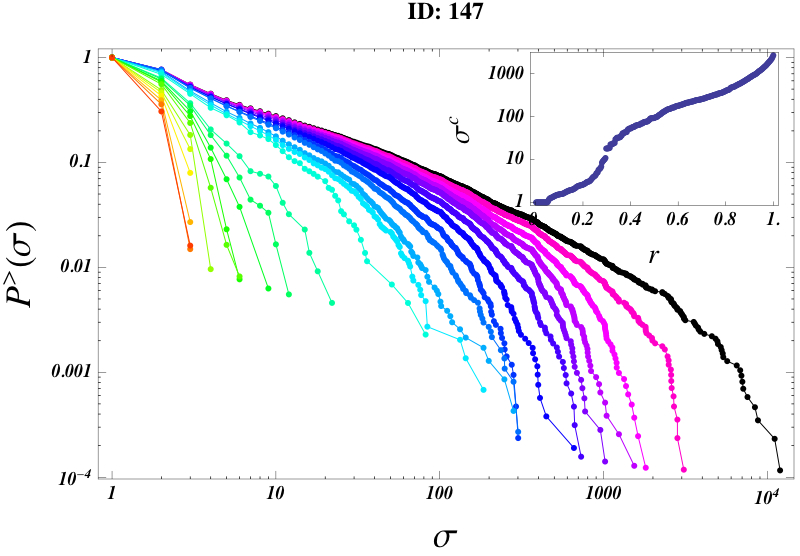} &\hspace{1mm} &  \includegraphics[width=40mm,clip=]{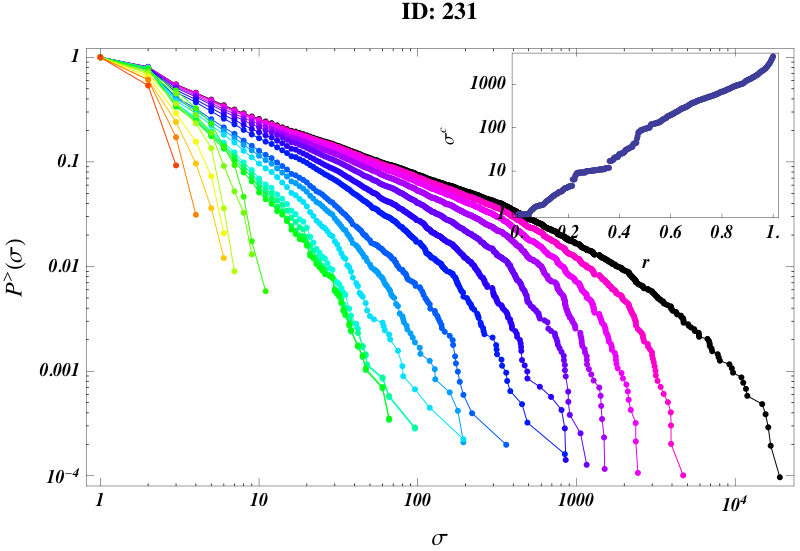}\\
\includegraphics[width=40mm,clip=]{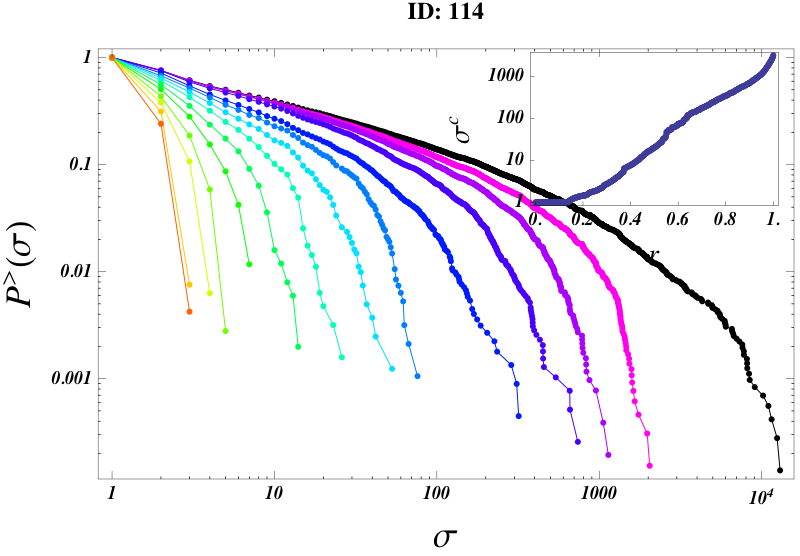} &\hspace{1mm} &  \includegraphics[width=40mm,clip=]{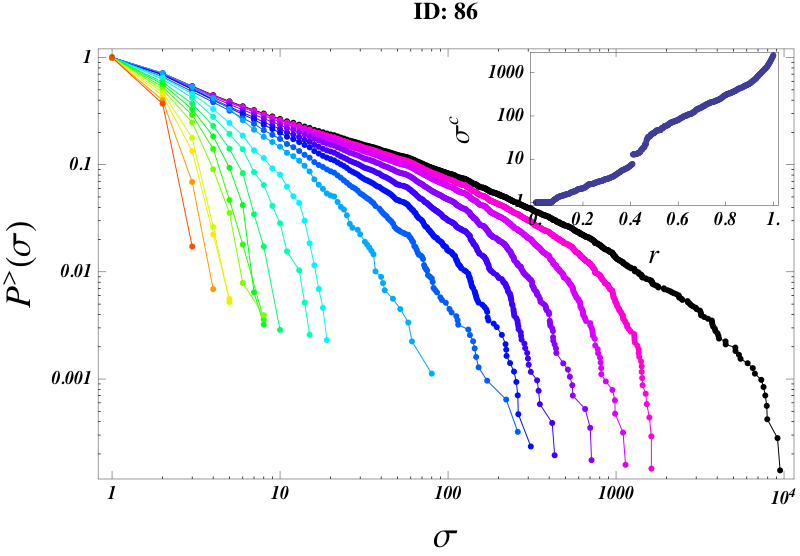}\\
\end{tabular}

\caption{\footnotesize Universal behavioral pattern of human agents. Log-log plots of the empirical response-time (RT) cumulative distributions  $P_{i}^{>}(\sigma)$ in \eqref{friends}, for six typical email writers in the database. The distributions $P_{i}^{>}(\sigma)$ are plotted following the same criteria as in Fig.~\ref{one_agent}. We see that for all agents the characteristic times $\sigma_{i}^{c}$ (shown in the insets) grow roughly exponentially with the normalized rank $r \uparrow 1$, see also Fig.~\ref{typical}. See the \href{https://www.researchgate.net/publication/280645729_Supplementary_figure_for_New_activity_pattern_in_human_interactive_dynamics}{Supplementary Figure} for more statistics.
}
\label{friends_empirical}
\end{figure}

For any given $i$, in \eqref{friends} we denote by $P^{>}_{i}(\sigma)$ the (inverse) cumulative RT distribution associated to the activity of $A$ with all her correspondents $C_j$ with $j<i$, i.e.~$P^{>}_{i}(\sigma)$ gives the probability of finding RTs longer than $\sigma$ when considering the responses of $A$ to any $C_j$ with $j<i$. In particular: $ P^{>}_{1} (\sigma)$ is the distribution of RTs of $A$ with her corresponent $C_1$; $ P^{>}_{2} (\sigma)$ is the distribution of the aggregated RTs of $A$ with her correspondents $C_1$ and $C_2$; and so on, with $ P^{>}_{n} (\sigma) \equiv P^{>}(\sigma)$ giving the total cumulative RT distribution of agent $A$ with all her correspondents. Fig.~\ref{one_agent} shows the empirical RT distributions $P^{>}_{i}(\sigma)$ belonging to the family $\mathcal{F}_{A}$ of a typical agent $A$ in the database. A quantitative description of the overall features of a family $\mathcal{F}_{A}$ is obtained by computing the characteristic time $\sigma_{i}^{c}$ pertaining to each $P^>_{i}(\sigma) \in \mathcal{F}_{A}$, i.e.~the values $\sigma_{i}^{c} = \frac{<\sigma^2>}{<\sigma>}$~ computed for the RTs of $A$ with all her correspondents $C_j$, with $j<i$. The $\sigma_{i}^{c}$ indicate how the $s$-clocked RTs associated to the correspondents of $A$ up to the $i$-th rank, grow longer as a whole with $i$. These values, which grow monotonically with $i$, measure how rapidly the curves $P^{>}_{i}(\sigma)$ progressively spread apart from each other on the plane $(\sigma, P)$ for growing $\sigma$, as they approach, for $i \uparrow n$, the total cumulative distribution $P^{>}(\sigma) \equiv P^{>}_{n}(\sigma)$ of $A$ (this is the upper-most curve, obtained for $i = n$, shown in black in each panel of Fig.~\ref{friends_empirical}). We find in Fig.~\ref{one_agent} that in the family $\mathcal{F}_{A}$ the individual distributions $P^>_{i}(\sigma)$ are heavy-tailed, but they are not power laws for $i<n$, and do not warrant any simple fitting form nor collapse property.  We also observe in the inset of Fig.~\ref{one_agent} that the characteristic times $\sigma_{i}^{c} $ computed for each distribution $P^{>}_{i}(\sigma)$, grow roughly exponentially with rank $i \uparrow n$ (the inset of Fig.~\ref{one_agent} shows $\sigma_{i}^{c}$ as a function of the normalized correspondent rank $r = \frac{i}{n} \in [0, 1]$).

As mentioned, the analysis in Ref.~[13] has revealed a strong universality in written communication, showing that the activity-clocked total RT probability densities $P(\sigma)$ of correspondence writers have the form of exponentially truncated power laws, with empirical exponents averaging near $-\frac{3}{2}$ across all correspondence media (letters, email, text messaging). An even stronger form of behavioral universality emerges from the analysis of the response patterns of humans described by the empirical distribution families $\mathcal{F}_{A} = \{P^>_{i}(\sigma)\}$ in \eqref{friends}. Indeed, Fig.~\ref{friends_empirical} shows the $\mathcal{F}_{A}$ relative to six typical active writers $A$, where we see how the distribution families pertaining to different agents exhibit largely the same features as those evidenced in Fig.~\ref{one_agent}, clearly pointing to a common pattern in these agents' reactive dynamics. More statistics of this type are given in the \href{https://www.researchgate.net/publication/280645729_Supplementary_figure_for_New_activity_pattern_in_human_interactive_dynamics}{Supplementary Figure}, wherein the common pattern in the families $\mathcal{F}_{A}$ highlighted in Figs.~\ref{one_agent}-\ref{friends_empirical} is not recognizable only for the least active agents in the database.  

\begin{figure}[t!]\centering
\begin{tabular}{ccc}
\includegraphics[width=41mm,clip=]{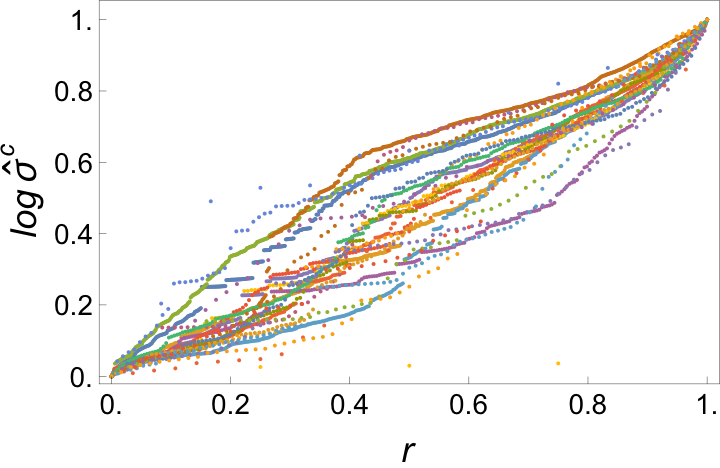} &\hspace{1mm} &
\includegraphics[width=42mm,clip=]{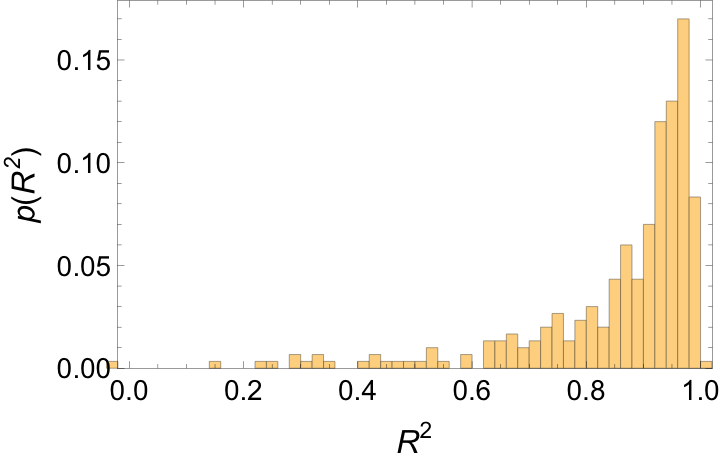} \\
(a) &  & (b)
\end{tabular}
\caption{\footnotesize Behavior of the characteristic times in the empirical data. (a)  Linear-log plot of the normalized characteristic times $\hat\sigma^{c} \in [0, 1]$, vs.~the normalized correspondent rank $r = \frac{i}{n} \in [0, 1]$, for a random sample of $\sim$30 agents among the 300 in the database. 
(b) Relative frequencies of the $R^2$ values obtained by fitting the empirical $\hat\sigma^{c}$-curves with the normalized exponential function $\exp(r)$, for all 300 agents. About 87\% of them have $R^2 > 0.70$, with 59\% having $R^2 > 0.90$. The agents whose RT families are shown in Fig.~\ref{friends_empirical} have $R^2$ values in the range $0.87$--$0.97$.
}
\label{typical}
\end{figure}

To establish more precisely the statistical commonality indicating the universal behavior of email writers, we analyze the characteristic times $\sigma_{i}^{c}$ pertaining to different agents. Fig.~\ref{typical}(a) shows the behavior of the normalized characteristic times $\hat\sigma^{c} \in [0, 1]$ for a random sample of writers in the database (see more data in the \href{https://www.researchgate.net/publication/280645729_Supplementary_figure_for_New_activity_pattern_in_human_interactive_dynamics}{Supplementary Figure}). A metric can be considered on the set of characteristic-times curves by analyzing the behavior of the $R^2$ parameter, for all the 300 agents in the dataset. The histogram in Fig.~\ref{typical}(b) gives the relative frequency plot for the fit of such curves by a normalized exponential function. We see the histogram is strongly peaked, confirming that the $\sigma_{i}^{c}$ grow roughly exponentially with rank for the great majority of agents, and giving quantitative confirmation to the universality of the activity pattern of email writers revealed by Figs.~\ref{one_agent}-\ref{friends_empirical} and the\href{https://www.researchgate.net/publication/280645729_Supplementary_figure_for_New_activity_pattern_in_human_interactive_dynamics}{Supplementary Figure}. 

\section{Model:~prioritization}~
Previous work \cite{Barabasi:2005yq,Vazquez:2006zr, Oliveira:2005fk, Gabrielli2009, Gabrielli2009-1, Grinstein2006,Grinstein2008, IAT1, IAT2, UWC} has analyzed various aspects of priority queueing in relation to written correspondence. Here we show how a simple model based on prioritization as in Ref.~[13], 
which accurately describes the power-law behavior of the total $s$-clocked RT  distribution $P(\sigma) = P_{n}(\sigma)$ observed in correspondence writers, also accounts robustly for the RT patterns described by the family $\mathcal{F}_{A}$ of empirical $s$-clocked distributions $P^{>}_{i}(\sigma)$ as in Figs.~\ref{one_agent}-\ref{friends_empirical}. 

We describe the model in its simplest form, suitable for agents whose $P(\sigma)$ exponent is (close to) $-1.5$. For different individual exponents see Ref.~[13]. 
We consider for an agent $A$ an initial list of $L$ tasks, with assigned priorities $y$ sampled from the uniform distribution on $[0,1]$. At each time step (which corresponds to a unit increment of the activity parameter $s$) the task with highest priority in the list is executed (a message replied), and $m > 1$ new tasks are added on average  to the list, each one with priority $y$ sampled as above.  It was analytically proven \cite{Gabrielli2009,Gabrielli2009-1, Abate1996} that this queueing mechanism produces an RT probability density $P(\sigma)$ which for $s \rightarrow \infty$ decays as a power law with exponent $-\frac{3}{2}$. When finite values of $s$ are considered as in numerical simulations, a truncated ($-\frac{3}{2}$)-power-law $P(\sigma)$ is obtained for the RTs. Interestingly, we compute that also such finite-size effect obtained in the model agrees with the cut-off observed in the scaling statistics from the empirical data, because for $s \sim10^4$ activity cycles, both the model and data give a characteristic time $\sigma^{c} \sim10^3$ for the total cumulative RT distribution $P^>(\sigma)$.

\begin{figure}[t!]\centering
\begin{tabular}{ccc}
\includegraphics[width=42mm,clip=]{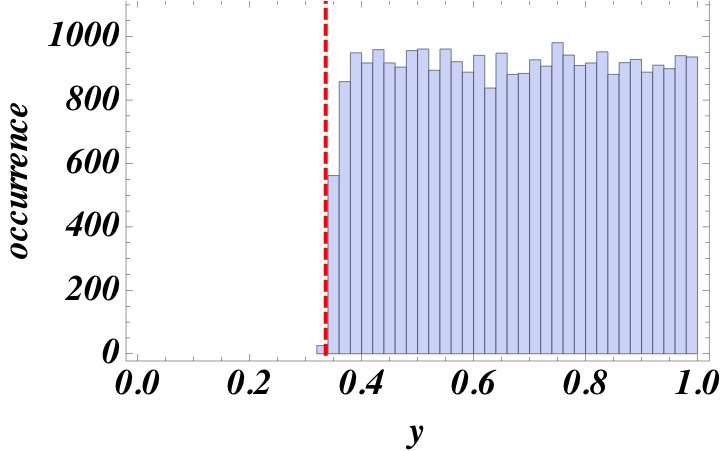} &\hspace{1mm} &
\includegraphics[width=42mm,clip=]{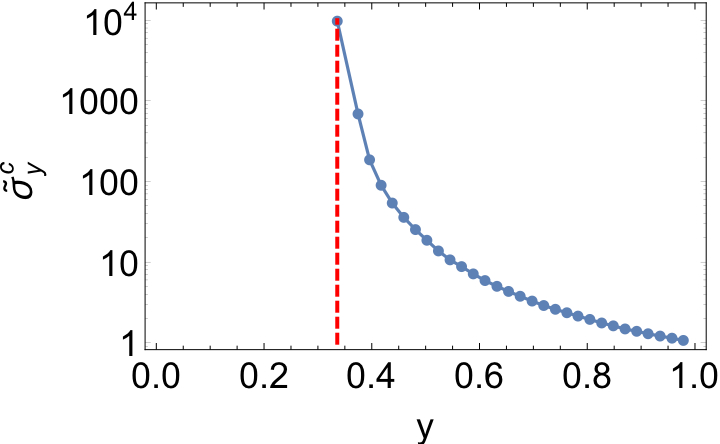} \\
(a) &  & (b)
\end{tabular}
\caption{\footnotesize Behavior of the priority model. (a) Histogram giving the priority $y$ of the replied-to messages in the model (for $3\cdot10^4$ cycles, with $\rho = m^{-1}  \simeq 0.64$), showing a threshold effect at the value $y \simeq 1-\rho$, marked by the vertical red line. (b) Linear-log plot for the corresponding characteristic times $\tilde\sigma_{y}^{c}$ generated by the model for $y \in [1-\rho, 1]$.
}
\label{threshold_model}
\end{figure}

Now, for the purpose of relating the model to the empirical features highlighted above regarding $\mathcal{F}_{A}$, it is natural to consider the priority-conditioned distributions that are generated by priority queueing. Specifically, let us consider the family of distributions $\mathcal{F} = \{P^>_y(\sigma)\}$, where $P^>_y(\sigma)$ is the probability of observing an RT larger than $\sigma$ given that the priority of the replied-to messages has values greater than $y$. When plotted, these $y$-conditioned distributions in $\mathcal{F}$ exhibit heavy tails, and, for decreasing $y$, fan out in the plane $(\sigma,P)$ in a way that is reminiscent of the empirical curves $P_i^>(\sigma)$ in Fig.~\ref{friends_empirical} for growing $i \uparrow n$. 
However, the distribution family $\mathcal{F}$ does not provide a good description for the families $\mathcal{F}_{A}$ obtained from the empirical data, because the characteristic times $\sigma_y^c=  \frac{<\sigma^2>}{<\sigma>}$ computed for the above $P^>_y(\sigma)$, i.e.~ by considering the aggregated RTs given by the model for all priorities greater than $y$, grow supra-exponentially as $y \downarrow 1-\rho$. This can be seen from the characteristic times $\tilde\sigma_{y}^{c}$ shown in Fig.~\ref{threshold_model}(b). Fig.~\ref{robust} below also shows the $\sigma_y^c$, whose supra-exponential theoretical values do not satisfactorily match the behavior of the empirical counterparts $\sigma_{i}^{c}$ for growing rank $i$, shown in the insets of Figs.~\ref{one_agent}-\ref{friends_empirical}.

\section{Gaussian averaging}~
To actually connect the model to the empirical families $\mathcal{F}_{A}$, we need to better understand the operation of the priority $y$ in generating the RTs in the queueing process. For this, we consider the distribution of $y$-values for the generated RTs, which are shown in Fig.~\ref{threshold_model}(a). We notice prioritization generates a threshold effect, whose existence can be proven in our context by adapting the arguments in~Ref.~[33] 
Accordingly, only the entering messages whose priority is $y \gtrsim 1-\rho$ are replied to in the model. As already mentioned, the characteristic times $\tilde\sigma_{y}^{c}$ of the corresponding RTs, shown in Fig.~\ref{threshold_model}(b), exhibit supra-exponential growth as the priority $y$ of the entering messages approaches the threshold value $1-\rho$ from above, and likewise behave the $\sigma_{y}^{c}$.

The failure of the family $\mathcal{F}$ to represent correctly the features of $\mathcal{F}_{A}$ is not surprising because, while certainly present, the correlation between the identity of correspondents and their messages' priority cannot be too strict, as each correspondent $C_i$ of $A$ should be associated, rather than to a single value of the priority $y$, to some individual distribution of $y$-values. To describe this, we consider a suitable family of kernels $\kappa(y; \bar{y}_i, d_i)$ which, for each $i = 1, 2, \dots$, describe in the model the distribution (with suitable s.d.~$d_i$, and with mean $\bar{y}_i$ decreasing with growing $i$) of priorities for the messages from the $i$-th correspondent $C_i$ which $A$ has replied to. Given such $\kappa$'s, we compute the distributions
\begin{equation}
\label{convolution}
P_{\bar{y}_i}(\sigma) =   \int P_y(\sigma)\kappa(y; \bar{y}_i, d_i)dy,
\end{equation}
where $P_y(\sigma)$ is the probability of observing, in the model, an RT in a small neighborhood of $\sigma$ given that the priority of the replied-to messages has values in a small neighborhood of $y$. Then, the behavior exhibited by the empirical families $\mathcal{F}_{A}$ in Fig.~\ref{friends_empirical} should be better captured by a new family of distributions 
\begin{equation}
\label{family_averaged}
\mathcal{\bar F} = \{P^>_{\bar{y}_i}(\sigma), \;\;  i = 1, 2, ... \, \}, 
\end{equation}
where the $P_{\bar{y}_i}^>(\sigma)$ are the (cumulative) distributions associated to the $P_{\bar{y}_i}(\sigma)$ in (\ref{convolution}), i.e., they give the probability of observing RTs greater than $\sigma$ when considering in the model the aggregated replied-to messages pertaining to all the $\kappa$-samples with average priorities $\bar{y}_j$ greater than $\bar{y}_i$.  

The simplest hypothesis in this context considers, for $i= 1, 2, \dots, n$, Gaussian kernels $\kappa(y; \bar{y}_i, d) = \frac{1}{d\sqrt{2\pi}} \exp\left(-\frac{1}{2}\left(\frac{y-\bar{y}_{i}}{d}\right)^2\right)$ in \eqref{convolution}, with a common s.d.~$d$, and values of the mean $\bar{y}_i$ which are homogeneously distributed between $1-\rho$ and 1 (these bounds derive from the threshold effect in the model, see Fig.~\ref{threshold_model}). The distribution of $y$-values over the RTs in the model remains largely homogeneous after sampling by means of these kernels, as in Fig.~\ref{threshold_model}(a). For $d \downarrow 0$ we recover the model with no averaging.

\begin{figure}[t!]\centering
 \includegraphics[width=70mm,clip=]{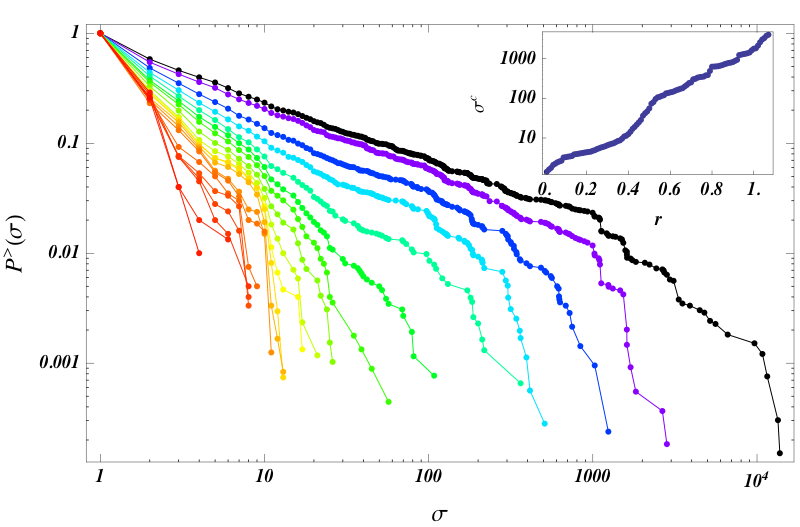} 
\caption{\footnotesize Behavior of the priority model with averaging. Log-log plots of the response-time (RT) inverse cumulative distributions $P_{\bar{y}_i}^{>}(\sigma)$ belonging to family $\mathcal{\bar F}$ in (\ref{family_averaged}), obtained through moving Gaussian averages of the $y$-conditioned probabilities computed from the model, for $3\cdot 10^4$ cycles, with $\rho \simeq 0.64$, $d=0.2$, and $n = 100$. Twenty curves are plotted, for geometrically growing values of the normalized rank $r = \frac{i}{n} \in [0, 1]$ from left to right (see text). The total RT distribution is the upper-most curve, shown in black, and $P^{>}_1(\sigma)$ is the lowest curve, in red. The inset shows the linear-log plot for the values $\sigma_{i}^{c}$ which increase roughly exponentially with rank (the horizontal axis in the inset reports the normalized rank $r$). See also Fig.~\ref{robust}. We observe the very good agreement with the behavior of the empirical families $\mathcal{F}_A$ in (\ref{friends}) which are shown in Figs.~\ref{one_agent}-\ref{friends_empirical} and in the \href{https://www.researchgate.net/publication/280645729_Supplementary_figure_for_New_activity_pattern_in_human_interactive_dynamics}{Supplementary Figure}.
}
\label{friends_computed}
\end{figure}

\begin{figure}[t]\centering

\begin{tabular}{ccc}
\includegraphics[width=41mm,clip=]{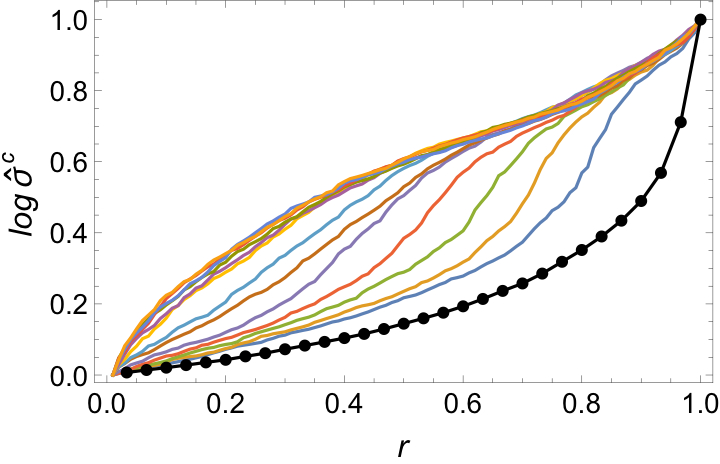} &\hspace{1mm} &
\includegraphics[width=42mm,clip=]{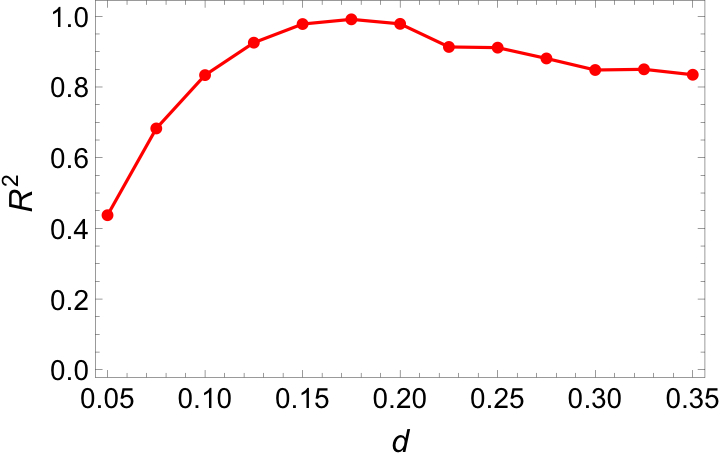} \\
(a) &  & (b)
\end{tabular}
\caption{\footnotesize Behavior of the characteristic times in the model. (a) Linear-log plot of the normalized characteristic times $\hat\sigma^{c} \in [0, 1]$ computed for a range of values $d \sim 10^{-1}$ for the s.d.~of the Gaussian kernels (colored curves). The dotted black curve gives the characteristic times for the case $d \downarrow 0$, i.e. for the model with no averaging. (b) Relation between $d$ value in the kernels and the $R^2$ values obtained by fitting the curves in panel (a) with the normalized exponential function $\exp(r)$. These high $R^2$ indicate the characteristic times obtained through Gaussian kernels with $d \sim 10^{-1}$ are well described by the exponential fit, unlike what happens in the model with no averaging for $d \downarrow 0$, as shown by the dotted curve in panel (a).
}
\label{robust}
\end{figure}

Fig.~\ref{friends_computed} shows the family of distributions $\mathcal{\bar F}$ in (\ref{family_averaged}) computed through the above Gaussian kernels, for $d=0.2$ and $n=100$. We see that the curves in the averaged distribution family $\mathcal{\bar F}$ do reproduce qualitatively very well the behavior of the empirical curves in the families $\mathcal{F}_{A}$ in Fig.~\ref{friends_empirical} and in the \href{http://iopscience.iop.org/1742-5468/2015/9/P09006/media/JSTAT_P09006_suppdata.pdf}{Supplementary Figure}. In particular, the inset in Fig.~\ref{friends_computed} shows that the values $\sigma_{i}^{c} =  \frac{<\sigma^2>}{<\sigma>}$, computed by aggregating the RTs given by the model referring to all the Gaussian samples with average priorities greater than $\bar{y}_i$, grow roughly exponentially with rank $i \uparrow n$ (i.e.~as $\bar{y}_i \downarrow 1-\rho$). This matches closely the behavior of their empirical $\sigma^{c}_{i}$ shown in the insets of Figs.~\ref{one_agent}-\ref{friends_empirical}. The analysis in Fig.~\ref{robust} corroborates the accord of $\mathcal{\bar F}$ with the empirical families $\mathcal{F}_{A}$, as we see that the computed characteristic times of $\mathcal{\bar F}$ (i.e.~the colored curves) in Fig.~\ref{robust}(a) behave like their empirical counterparts in Fig.~\ref{typical}(a) for a range of values $d$ of the order $10^{-1}$. This agreement of the averaged-model family $\mathcal{\bar F}$ with the empirical families $\mathcal{F}_{A}$ is assessed quantitatively in Fig.~\ref{typical}(b), which shows the high $R^2$ values, above 0.9, obtained for the exponential fit of the computed characteristic times for $\mathcal{\bar F}$ in this $d$-range. This indicates that the characteristic times in the averaged model robustly display roughly exponential growth as do their empirical counterparts in Figs.~\ref{friends_empirical}-\ref{typical}(b). This effect thus does not need fine tuning in the model, and is rather rooted in the prioritization process and the weighted averaging used to account for the priority distribution of the messages from each correspondent. 

\section{Discussion}
We have achieved in this study a two-fold result. (a) Firstly, we have uncovered a new universal activity pattern in the interactive dynamics of correspondence writers, highlighted through the examination of long-term empirical data on written correspondence via email. We find that agents all distribute in the same way their interactions separately with each one of their distinct contacts, generating families of heavy-tailed RT distributions which have largely the same features across writers, with characteristic times which universally exhibit roughly exponential growth with correspondent rank. This analysis considerably extends the scrutiny of the sole total distribution of correspondence writers, on which the literature has focused so far. (b) We have furthermore shown that this previously undetected behavioral pattern emerges robustly by considering Gaussian moving averages on the priority-conditioned RT probabilities derived from a basic priority model. 

Our findings clarify how priority-queueing contributes to generate the observed activity statistics of human response, and suggest the associated universal patterns may result from fundamental constraints imposed by prioritization and by averaging mechanisms on the outcome of any complex underlying individual choice processes. The effects reveled here should affect both the architecture and the evolution of communication and interaction (social) networks, imposing explicit constraints on their future exploration and modeling. They may also contribute to better estimate the possible value of such networks related to size, \cite{metcalf_wrong, network_exclusion} which is an important question in computer science, business management, and sociology. Natural extensions of the present study relate to the possibility of identifying, within the individual variations of the empirical $\sigma^{c}$-curves, the existence of core communities \cite{communities_dunbar, communities_plosone} within each agent's ensemble of correspondents. 
Another point of interest is the adoption of less schematic averaging kernels than used above. This would not affect the basic point of behavioral universality highlighted here, but may help capture other effects occurring in written communication, and in reactive dynamics in general. Indeed, we expect the stylized facts \cite{stylized}  and universal activity patterns presently uncovered for email correspondence can also occur, and could be successfully investigated, in other general interactive environments. This should  promote our understanding of the dynamics of reciprocal activity in diverse agent-driven domains, as in economics or sociology. For instance, our approach may enhance queueing-based models \cite{threshold, finance1, finance2} as valuable tools in finance for investigating order-book dynamics.
Also theories for preference formation and extraction, for competing-opinion dynamics, and for information spreading \cite{social1, social2, social3, social4}, may benefit from the knowledge and analysis of reciprocal-action data such as we have obtained here on emailing, because decision making at the personal and collective levels, or the shift of sentiments and preferences, are largely based on how individuals communicate and interact with each other. In general, the present analysis should help inform future empirical and theoretical work on the interplay among distinct agents of any kind, animate or inanimate, embedded in networks of reactive relations.

\section*{Acknowledgements}
We thank Dr.~M.~Gravino for providing to us the email data analyzed in this study. AM acknowledges the Cariparo Foundation for financial support. MF acknowledges financial support of  GA\v{C}R grant P201/12/2613. MF thanks prof. J.M.~Swart for many conversations.

\section*{Author contributions statement}
Authors with initials M.F., A.L., A.M. and G.Z. equally contributed to the manuscript. 

\section*{Additional information}
\textbf{Competing financial interests:} The authors declare no conflict of interest.

\end{document}